\input{aipcheck}

\documentclass[
    ,final            
  ]
  {aipproc}

\layoutstyle{6x9}

\begin{document}

\title{Cosmological Big Bounce Transition}


\author{W{\l}odzimierz Piechocki}{address={Theoretical Physics Department, Institute for
Nuclear Studies,  Ho\.{z}a 69, 00-681 Warsaw, Poland; e-mail:
piech@fuw.edu.pl}}

\begin{abstract}
We analyze the  big bounce transition of the quantum FRW model in
the setting of the nonstandard loop quantum cosmology (LQC).
Elementary observables are used to quantize compound observables.
The spectrum of the energy density operator is bounded and
continuous. The spectrum of the volume operator is bounded from
below and discrete. It has equally distant levels defining a
quantum of the volume. The discreteness may imply a foamy
structure of spacetime at semiclassical level which may be
detected in astro-cosmo observations. The nonstandard LQC method
has a free parameter that should be fixed in some way to specify
the big bounce transition.
\end{abstract}

\maketitle


\section{Introduction}

An evidence for the existence  of  the cosmological {\it
singularity} (diverging gravitational and matter fields
invariants) is roughly speaking the following: (i)  observational
cosmology - our universe emerged from a state with extremely high
energy densities of matter fields, and (ii) theoretical cosmology
- most of known general relativity (GR) models of the universe
(Bianchi, Friedmann, Lema\^{i}tre, Kasner, ...) predict an
existence of the cosmological singularities. An expectation is
that {\it quantization} may heal the singularities.

There are two alternative methods of canonical quantization of
cosmological models of  GR, which make use of the {\it loop}
geometry:
\begin{itemize}
      \item standard LQC, i.e.  Dirac's method, based on the rule:
      `first quantize then
      impose constraints';
      \item  nonstandard LQC, i.e. reduced phase space (RPS) quantization,
      based on the rule: `first solve constraints then quantize'.
\end{itemize}
The results for flat FRW model with massless scalar field, are
roughly speaking, the following:
\begin{itemize}
    \item standard LQC: classical Big Bang is replaced by quantum
    Big Bounce due  to strong {\it quantum} effects at the Planck
    scale (see, e.g.
    \cite{Ashtekar:2006rx,Ashtekar:2006wn});
    \item  nonstandard LQC (see, e.g. \cite{Dzierzak:2009ip,Malkiewicz:2009qv}):
    \begin{itemize}
    \item  {\it modification} of GR by loop geometry is responsible
    for the  resolution of the singularity,
    \item quantization may lead to {\it discrete} spectra of physical
    observables, which may be confronted with the data of observational
    cosmology.
    \end{itemize}
\end{itemize}

There are many intriguing questions to be answered within quantum
cosmology. For instance:  What is the energy scale of the Big
Bounce? What is the structure of spacetime at semi-classical level
(simply connected, foamy or discrete)? What is the  origin of tiny
fluctuations in the energy density visible in the CMB? How  long
had the quantum phase lasted? What was before the Big Bounce?

In what follows we present results concerning quantization of the
two models: \\(i) isotropic and homogeneous universe - flat FRW
model with massless scalar field
\cite{Dzierzak:2009ip,Malkiewicz:2009qv}, and with cosmological
constant $\Lambda$ \cite{Mielczarek:2010rq,Mielczarek:2010wu}, and
(ii) homogeneous universe - the Bianchi I model with massless
scalar field \cite{Dzierzak:2009dj,Malkiewicz:2010py}.

\section{Classical Level}

\subsection{Modified Hamiltonian}

The gravitational part of  of the Hamiltonian of FRW universe with
massless scalar field reads \cite{Dzierzak:2009ip}
\begin{equation}\label{hamG}
H_g = - \gamma^{-2} \int_{\mathcal V} d^3 x ~
e^{-1}\varepsilon_{ijk}
 E^{aj}E^{bk}  F^i_{ab}\, ,
\end{equation}
where  $~\gamma$, Barbero-Immirzi parameter; $\mathcal V\subset
\Sigma$, fiducial cell; $\varepsilon_{ijk}$, alternating tensor;
$E^a_i $, density weighted  triad; $e:=\sqrt{|\det E|}$; $
F^k_{ab} =
\partial_a A^k_b - \partial_b A^k_a + \epsilon^k_{ij} A^i_a
A^j_b$, curvature of $SU(2)$  connection $A^i_a$.

By {\it modification} we mean the following approximation applied
to $ F^k_{ab}:$
\begin{equation}\label{finite}
F^k_{ab}(\lambda) \approx
-2\;Tr\;\Big(\frac{h^{(\lambda)}_{L_{ij}}-1}{\lambda^2
}\Big)\;{\tau^k}\; ^o\omega^i_a  \; ^o\omega^j_a .
\end{equation}
where holonomy of the connection, around the square (loop) $
L_{ij}$ with sides length $\lambda $, is $~h^{(\lambda)}_{L_{ij}}
= h^{(\lambda)}_i h^{(\lambda)}_j (h^{(\lambda)}_i)^{-1}
(h^{(\lambda)}_j)^{-1}$. An exact formula reads
\begin{equation}\label{exa}
F^k_{ab}= \lim_{\lambda\,\rightarrow \,0}\, F^k_{ab}(\lambda).
\end{equation}

The holonomy $ h_s(A)$ of connection $A$ along a curve $s:
[0,1]\rightarrow s(t)\in \mathcal{V}$ is the solution to the
equation
\begin{equation}\label{holeq}
      \frac{d}{d t} h_{s_t}(A)= A(s_t) h_{s_t}(A),
     ~~~~h_{s_0}= I,
\end{equation}
where $A(s_t):= A^j_a (s_t)\,\tau_j\,\dot{s}^a (t),~~
h_s(A):=h_{s_1}(A)\in SU(2),~~s_t \equiv s(t).$

The holonomy  along straight edge $ ^oe^a_k\partial_a $ of length
$\lambda $ (in fundamental, j = 1/2, representation of SU(2)
group) reads
\begin{equation}\label{ho1}
 h^{(\lambda)}_k (c) =  \exp (\tau_{k}\lambda c) = \cos (\lambda
c/2)\;I + 2\,\sin (\lambda c/2)\;\tau_k,
\end{equation}
where $A^k_a = \,^o\omega^k_a\,c \,$ and $\tau_k = -i
\sigma_k/2\;$ ($\sigma_k$ are the Pauli spin matrices).

The total Hamiltonian for FRW universe with a massless scalar
field $\phi$ is given by
\begin{equation}\label{ham}
   H = H_g + H_\phi,~~~~H_\phi := p^2_\phi \,
|p|^{-\frac{3}{2}}/2
\end{equation}
where $\phi$ and $p_\phi$ are elementary variables satisfying
$\{\phi,p_\phi\} = 1$.

Making use of (\ref{ho1}) we calculate  (\ref{hamG}) and get the
modified  total Hamiltonian  corresponding to (\ref{ham})
\begin{equation}\label{regH}
   H^{(\lambda)}/N = -\frac{3}{8\pi G \gamma^2}\;\frac{\sin^2(\lambda
\beta)}{\lambda^2}\;v + \frac{p_{\phi}^2}{2\, v},
\end{equation}
where the  canonical variables (of the improved scheme) read
\begin{equation}\label{re1}
    \beta := \frac{c}{|p|^{1/2}},~~~v := |p|^{3/2} ,
\end{equation}
where $\beta\sim \dot{a}/a~$ and $v \sim a^3\,$ (for $\lambda
=0$). Eq (\ref{regH}) presents a modified {\it classical}
Hamiltonian; it includes no quantum physics.

The Poisson  bracket for the canonical variables
$(\beta,v,\phi,p_\phi)$ is defined to be
\begin{equation}\label{re2}
    \{\cdot,\cdot\}:= 4\pi G\gamma\;\bigg[ \frac{\partial \cdot}
    {\partial \beta} \frac{\partial \cdot}{\partial v} -
     \frac{\partial \cdot}{\partial v} \frac{\partial \cdot}{\partial \beta}\bigg] +
     \frac{\partial \cdot}{\partial \phi} \frac{\partial \cdot}{\partial p_\phi} -
     \frac{\partial \cdot}{\partial p_\phi} \frac{\partial \cdot}{\partial
     \phi}.
\end{equation}
The {\it dynamics} of a canonical variable $\xi$ is given by
Hamilton's equations
\begin{equation}\label{dyn}
    \dot{\xi} := \{\xi,H^{(\lambda)}\},~~~~~~\xi \in \{\beta,v,\phi,p_\phi\},
\end{equation}
where $\dot{\xi} := d\xi/d\tau$, and where $\tau$ is an evolution
parameter.  Dynamics in   {\it physical} phase space,
$\mathcal{F}_{phys}^{(\lambda)}$, is defined by solutions to
(\ref{dyn}) satisfying the condition $H^{(\lambda)}\approx 0$.
Solutions of (\ref{dyn}) ignoring the constraint
$H^{(\lambda)}\approx 0$ are in    {\it kinematical} phase space,
$\mathcal{F}_{kin}^{(\lambda)}$.

\subsection{Physical phase space}

Equation (\ref{regH}) can be rewritten as
\begin{equation}\label{product}
  H^{(\lambda)} = N\,H_0^{(\lambda)}\,\tilde{H}^{(\lambda)}\approx 0,
\end{equation}
where
\begin{equation}\label{defprod}
H_0^{(\lambda)} := \frac{3}{8 \pi G \gamma^2 v} \;\Big(\kappa
\gamma |p_\phi| + v\,\frac{|\sin(\lambda
\beta)|}{\lambda}\Big),~~~~~~ \tilde{H}^{(\lambda)}:= \kappa
\gamma |p_\phi| - v\, \frac{|\sin(\lambda \beta)|}{\lambda}.
\end{equation}
It is clear that $H_0^{(\lambda)} = 0$  iff $p_\phi
=0=\sin(\lambda \beta)$. In such a case $\tilde{H}^{(\lambda)}
=0$, thus $H^{(\lambda)}$ equals identically zero so there is no
dynamics. We exclude such pathological case from further
considerations. Since $N\,H_0^{(\lambda)}\neq 0$, the original
dynamics may be reduced to the  relative dynamics with the simpler
constraint, $\tilde{H}^{(\lambda)}\approx 0$.

For functions $f$ and $g$ on $\mathcal{F}_{phys}^{(\lambda)}$ we
have
\begin{equation}\label{funF}
    \dot{f} = \{f, N H_0^{(\lambda)}\tilde{H}^{(\lambda)}\} =
     N H_0^{(\lambda)}\{f,\tilde{H}^{(\lambda)}\},
    \end{equation}
\begin{equation}\label{funG}
    \dot{g} = \{g, N H_0^{(\lambda)}\tilde{H}^{(\lambda)}\} =
    N H_0^{(\lambda)}\{g,\tilde{H}^{(\lambda)}\}.
\end{equation}
The relation
\begin{equation}\label{fraC}
    \frac{\dot{f}}{\dot{g}} = \frac{df}{dg} = \frac{N H_0^{(\lambda)}
    \{f,\tilde{H}^{(\lambda)}\}}
    {N H_0^{(\lambda)}\{g,\tilde{H}^{(\lambda)}\}} = \frac{\{f,\tilde{H}^{(\lambda)}\}}
    {\{g,\tilde{H}^{(\lambda)}\}}
    ,~~~~\mbox{as}~~~~H_0^{(\lambda)} \neq 0,
\end{equation}
may be rewritten as
\begin{equation}\label{integ}
     \frac{df}{\{f,\tilde{H}^{(\lambda)}\}} = \frac{dg}
    {\{g,\tilde{H}^{(\lambda)}\}} .
\end{equation}
Thus, in the relative dynamics one canonical variable may be used
as an `evolution parameter' if that variable is a {\it monotonic}
function.

Equations of motion, in the gauge $N = 1/H_0^{(\lambda)}$,  read
\begin{equation}\label{1a}
\dot{\phi} =
\kappa\gamma~\textrm{sgn}(p_{\phi}),~~~~~\dot{p_{\phi}}=0,
\end{equation}
\begin{equation}\label{2a}
\dot{\beta}=  -4\pi G\gamma \;\frac{|\sin(\lambda\,
\beta)|}{\lambda},~~~
  \dot{v} = 4\pi G\gamma v \cos(\lambda\,
  \beta)~\textrm{sgn}(\sin(\lambda\, \beta)),
\end{equation}
\begin{equation}\label{3a}
\tilde{H}^{(\lambda)}  \approx 0,
\end{equation}
where $\kappa^2 = 4\pi G/3.$ Due to (\ref{1a}), $ \phi$ is a
monotonic function.

The solution of the relative dynamics is found to be:
\begin{equation}\label{res2}
v = \frac{\Delta}{2}\cosh\Big(3\kappa\,s\,(\phi - \phi_{0})-\ln
\Delta \Big),~~~~\beta = \frac{1}{\lambda}  \arcsin (\Delta/v) ,
\end{equation}
where $\Delta:=\kappa\gamma\lambda|p_\phi|$ and
$s:=\textrm{sgn}(p_{\phi})$.  The variables $ v$ and $ \beta$ are
functions of the evolution parameter $ \phi$.

\subsection{Elementary observables}

A function, $\mathcal{O}: \mathcal{F}_{kin}^{(\lambda)}\rightarrow
R$, is a Dirac observable  if
\begin{equation}\label{dirac}
\{\mathcal{O},H^{(\lambda)}\} = 0.
\end{equation}
Thus, $\mathcal{O}$ is solution to the equation
\begin{equation}\label{dir}
\frac{\sin(\lambda\beta)}{\lambda}\,\frac{\partial
\mathcal{O}}{\partial\beta} - v \cos(\lambda\beta)\,\frac{\partial
\mathcal{O}}{\partial v} -
\frac{\kappa\gamma\,\textrm{sgn}(p_{\phi})}{4 \pi
G}\,\frac{\partial \mathcal{O}}{\partial\phi} = 0.
\end{equation}
The simplest solutions to (\ref{dir}), which we call {\it
elementary} observables, are found to be
\begin{equation}\label{obser1}
\mathcal{O}_1:= p_{\phi},~~~\mathcal{O}_2:= \phi -
\frac{\textrm{sgn}(p_{\phi})}{3\kappa}\;\textrm{arth}\big(\cos(\lambda
\beta)\big),~~~~ \mathcal{O}_3:= \textrm{sgn}(p_{\phi})\,v\,
\frac{\sin(\lambda \beta)}{\lambda}.
\end{equation}
One may easily verify that the elementary observables satisfy the
Lie algebra
\begin{equation}\label{ala1}
\{\mathcal{O}_2,\mathcal{O}_1\}=
1,~~~~\{\mathcal{O}_1,\mathcal{O}_3\}= 0,~~~~
\{\mathcal{O}_2,\mathcal{O}_3\}=  \gamma\kappa .
\end{equation}

Due to the constraint $\tilde{H}^{(\lambda)}=0$, we have
\begin{equation}\label{con}
\mathcal{O}_3=  \gamma \kappa \,\mathcal{O}_1.
\end{equation}

One can show that in the physical phase space,
$\mathcal{F}_{phys}^{(\lambda)}$, we have only two elementary
observables which satisfy the algebra
\begin{equation}\label{alg1}
 \{\mathcal{O}_2,\mathcal{O}_1\}= 1,
\end{equation}
where
\begin{equation}\label{sym3}
\{\cdot,\cdot\}:=\frac{\partial\cdot}{\partial
\mathcal{O}_2}\frac{\partial\cdot}{\partial \mathcal{O}_1} -
\frac{\partial\cdot}{\partial
\mathcal{O}_1}\frac{\partial\cdot}{\partial \mathcal{O}_2}.
\end{equation}

The space $\mathcal{F}_{kin}^{(\lambda)}$ is  four dimensional. In
relative dynamics one variable is used to parametrize three
others. Since the constraint relates two variables, we have only
two independent variables.

\subsection{Compoud observables}

By {\it compound} observables we mean functions on phase space
which can be expressed in terms of elementary observables and an
evolution parameter $ \phi$,  so they are not observables.
However, they do become observables for each {\it fixed} value of
$ \phi$, since in such a case they are only functions of
elementary observables.

In what follows we consider two compound observables: the {\it
energy} density of matter field
\begin{equation}\label{rho2}
\rho(\phi,\lambda) =
\frac{1}{2}\,\frac{1}{(\kappa\gamma\lambda)^2\,
    \cosh^2 3\kappa  (\phi- \mathcal{O}_2)},
\end{equation}
and the {\it volume} operator
\begin{equation}\label{vol}
    v(\phi,\lambda) = \kappa\gamma\lambda\,
    |\mathcal{O}_1|\,\cosh3\kappa  (\phi-
    \mathcal{O}_2).
\end{equation}

\section{Quantum Level}

\subsection{Energy density operator}

In the Schr\"{o}dinger representation we have
\cite{Malkiewicz:2009qv}
\begin{equation}\label{quant1}
    \mathcal{O}_1 \rightarrow \widehat{\mathcal{O}}_1 :=
\widehat{x}  := x ,~~~~\mathcal{O}_2 \rightarrow
\widehat{\mathcal{O}}_2 := -i\,\hbar\,\partial_x ,
\end{equation}
so the energy density operator reads
\begin{equation}\label{quant3}
    \widehat{\rho}:=\frac{1}
    {2(\kappa\gamma\lambda)^2\cosh^2 3\kappa(\phi+i\,\hbar\,\partial_x)}.
\end{equation}
Solution to the eigenvalue problem
\begin{equation}\label{quant4}
\widehat{\rho}\,f_p = \rho (p)\,f_p ,
\end{equation}
reads
\begin{equation}\label{sol1}
f_p (x) = (2\pi)^{-1/2} \exp(i x p /\hbar),~~~~ \rho (p) =
\frac{1}{2}\frac{1}
    {(\kappa\gamma \lambda)^2\cosh^2 3\kappa(\phi - p)},
\end{equation}
where $p \in R$. So the energy density has the same functional
form as the classical one.

The density $\rho$ has  maximum  at the minimum of $v$:
\begin{equation}\label{cr1}
\rho_{\max} = \frac{1}{2\kappa^2 \gamma^2}\,\frac{1}{ \lambda^2}.
\end{equation}
For the  Planck scale, substituting $ \lambda = l_{Pl}$  into
(\ref{cr1}) gives $ \rho_{max}/\rho_{Pl} \simeq 2,07$; for $
\rho_{\max} = \rho_{Pl}$ we get $ \lambda \simeq 1,44\;l_{Pl}$.
Thus, Eg. (\ref{cr1})  fits the Planck scale.

The spectrum $(0,\frac{1}{2(\kappa \gamma \lambda)^2})$ is {\it
continuous} and  {\it bounded} for  $\lambda\neq 0$. For $\lambda
\rightarrow 0$ we get the classical FRW singularity. Since
$\lambda$ is a {\it free} parameter, finding the critical density
of matter corresponding to the Big Bounce is an open problem.

\subsection{Volume operator}

The classical volume operator, $v$, can be presented as
\begin{equation}\label{vvol}
v = |w|,~~~w :=
\kappa\gamma\lambda\;\mathcal{O}_1\;\cosh3\kappa(\phi-
\mathcal{O}_2).
\end{equation}
For  $\mathcal{O}_1$ and $\mathcal{O}_2$ we use the
Schr\"{o}dinger representation:
\begin{equation}\label{rep1}
\mathcal{O}_1 \longrightarrow \widehat{\mathcal{O}}_1 f(x):=
-i\,\hbar\,\partial_x f(x),~~~~ \mathcal{O}_2 \longrightarrow
\widehat{\mathcal{O}}_2 f(x):= \widehat{x} f(x) := x f(x).
\end{equation}
Thus, an explicit form of $\hat{w}$ is given by
\begin{equation}\label{repp1}
 \hat{w}= i\,\frac{\kappa\gamma\lambda\hbar}{2}\Big(
    2 \cosh3\kappa(\phi-x)\;\frac{d}{dx}
     -3\kappa\sinh3\kappa
    (\phi-x)\Big).
\end{equation}

Solution to the eigenvalue problem is found to be:
\begin{equation}\label{eq4}
\hat{w}\, f_a (x) = a\,f_a (x),~~~a \in R ,
\end{equation}
\begin{equation}\label{eq5}
f_a (x):= \frac{\sqrt{\frac{3\kappa}{\pi}}\exp\big(i \frac{2
a}{3\kappa^2 \gamma\lambda\hbar}\arctan
    e^{3\kappa(\phi-x)}\big)}{\cosh^{\frac{1}{2}}3\kappa(\phi-x)},
\end{equation}
\begin{equation}\label{eqq5}
     a = b + 6\kappa^2\gamma\lambda\hbar\, m =  b + 8\pi
    G\gamma\lambda\hbar\, m,
\end{equation}
where $b \in R$ and $ m\in Z$. Completion of the span of
\begin{equation}\label{set1}
\mathcal{F}_b:=\{~f_a\;|\; a = b + 8\pi G\gamma\lambda\hbar\,
m\}\subset L^2(R),
\end{equation}
in the norm of $L^2(R)$ leads to $L^2(R)$, $\forall b \in R$. The
operator $\hat{w}$ is essentially {\it self-adjoint} on each span
of $\mathcal{F}_b$.

Due to the the relation (\ref{vvol}), and the spectral theorem on
self-adjoint operators, we get the solution of the eigenvalue of
the volume operator:
\begin{equation}\label{sp1}
 v = |w|~~~\longrightarrow~~~\hat{v} f_a :=  |a| f_a .
\end{equation}
The spectrum is {\it bounded} from below and {\it discrete.} There
exists the minimum gap $ \bigtriangleup := 8\pi
G\gamma\hbar\,\lambda\;$ in the spectrum, which defines a {\it
quantum} of the volume. In the limit $ \lambda \rightarrow 0$,
corresponding to the   classical  FRW model, there is no quantum
of the volume.

\section{Extension of results}

\subsection{FRW model with cosmological constant}

The Hamiltonian of the flat  FRW model with a free massless scalar
field and the cosmological constant $ \Lambda >0$ has the form
\cite{Mielczarek:2010rq,Mielczarek:2010wu}:
\begin{equation}
H^{(\lambda)} = -\frac{3}{8\pi G \gamma^2} \frac{\sin^2(\lambda
\beta) }{\lambda^2} v +\frac{p^2_{\phi}}{2 v} +v \frac{
\Lambda}{8\pi G} .
\end{equation}
The eigenvalue problem for the  volume operator has the solution
\begin{equation}\label{e1}
    \hat{v} \psi_b = |b|\, \psi_b,~~~\psi_b \in L^2 (R),
\end{equation}
\begin{equation}\label{ev1}
b =  a + \frac{4\pi^2 G\hbar \gamma  \lambda m}
{\arctan\sqrt{1/\delta -1}}, ~~~~~m =0,1,2,\ldots
\end{equation}
where $0 \neq a \in R$ and $\delta := \Lambda \gamma^2 \lambda^2
/3$. The spectrum is {\it discrete}.

\subsection{Bianchi I model}

In the case of the Bianchi I model with free massless scalar field
we have \cite{Dzierzak:2009dj}
\begin{equation}\nonumber
H^{(\lambda)}= -\frac{1}{8\pi G
\gamma^2\lambda^2}\;\bigg[|p_1p_2|^{3/2}\sin(c_1 \mu_1)\sin(c_2
\mu_2) + \textrm{cyclic} \bigg] + \frac{p_{\phi}^2}{2},
\end{equation}
where $\mu_k:= \sqrt{\lambda /|p_k|}$, for $k=1,2,3.$

The solution to the eigenvalue problem for the directional $k$-th
volume operator (k = 1,2,3), for the Kasner-unlike case, reads
\cite{Malkiewicz:2010py}
\begin{equation}\nonumber
    \hat{v}_\alpha \psi_\alpha = \alpha\, \psi_\alpha,~~~\psi_\alpha \in
    L^2 ([0,\pi/b]),~~~\alpha = ab, 2ab, 3ab, \ldots
\end{equation}
where $a = \kappa \gamma \lambda,~~b = 3 \kappa,~~ab = 4\pi G
\gamma\lambda\hbar$.

In the Kasner-like case, two directional volume operators $w_2$
and $w_3$ have the spectra similar to the Kasner-unlike case. The
volume $w_1$ has the spectrum similar to the spectrum of a
harmonic oscilator in a box \cite{Malkiewicz:2010py}.

In all cases the spectrum is {\it discrete} and does not include
the {\it zero} volume eigenvalue.

\section{Summary}

Conclusions:
\begin{itemize}
      \item {\it Modification} of classical
       Hamiltonian, realized by making use of the loop geometry,
       turns big bang into big bounce (BB).
      \item No specific $0< \lambda \in R$ $~~\Rightarrow~~$  no
      specific {\it energy density} at BB
   \begin{itemize}
      \item BB may occur at {\it any} low density, e.g. at the density of water
        - disagreement with  observations;
      \item BB may occur at {\it any} high density, e.g. at $10^{500}$ of the
      Planck density
      - conflict with common believe;
      \item assuming $\; \lambda =
       l_{Pl}\;$ leads to $\; \rho_{max}/\rho_{Pl} \simeq 2,07$,
       so our model fits the Planck scale.
      \end{itemize}
  \item {\it Discreteness} of the spectra of the volume
  operators  may favor a {\it foamy} structure of space at short
        distances that may be detected in astro-cosmo
        observations.
\end{itemize}

Determination of the {\it free} parameter $\lambda$:
\begin{itemize}
      \item Loop Quantum Cosmology
      \begin{itemize}
      \item FRW model: no $\lambda$  is  privileged,
      \item Bianchi I model: no $\lambda$  is privileged.
      \end{itemize}
      \item Observational Cosmology
      \begin{itemize}
      \item   cosmic {\it photons}: no dispersion of photons \cite{Aharonian:2008kz}
      up to the energy $5\times 10^{17}$ GeV,
      \item detection of primordial  gravity {\it waves:} we hope that imprints of tensor modes
      on CMB spectrum \cite{Mielczarek:2008pf} may be used to
      determine $\lambda$.
\end{itemize}
\end{itemize}

\section{Challenge}

A great challenge is quantization of the
Belinskii-Khalatnikov-Lifshitz scenario (see
\cite{Belinski:2009wj} and references therein):
\begin{itemize}
      \item  general solution of GR (corresponding to a non-zero measure subset
      of all initial conditions);
       \item presents an evolution of spacetime near the space-like
       singularity (with diverging gravitational and matter fields
       invariants);
       \item singular solution applies both to the future singularity
       (Big Crunch) and past singularity (Big Bang);
       \item  BKL scheme appears in the low energy limit of superstring
       models (with Kac-Moody algebras as an underlying
       mathematics);
       \item unclear if loop LQG/LQC can be used to
       resolve the singularity problem of BKL.
\end{itemize}

The quantum BKL scenario (after being constructed) may give the
answer to the questions mentioned in the Introduction.

\begin{theacknowledgments}
The author would like to thank the organizers for inspiring
atmosphere at the Meeting.
\end{theacknowledgments}



\bibliographystyle{aipproc}   


\IfFileExists{\jobname.bbl}{}
 {\typeout{}
  \typeout{******************************************}
  \typeout{** Please run "bibtex \jobname" to optain}
  \typeout{** the bibliography and then re-run LaTeX}
  \typeout{** twice to fix the references!}
  \typeout{******************************************}
  \typeout{}
 }



\end{document}